\documentclass[letterpaper, 10 pt, conference]{ieeeconf}  

\IEEEoverridecommandlockouts                              

\usepackage{graphics} 
\usepackage{epsfig} 
\usepackage{amsmath} 
\usepackage{amssymb}  
\usepackage[caption=false]{subfig}

\newtheorem{definition}{Definition}
\newtheorem{theorem}{Theorem}
\newtheorem{lemma}{Lemma}
\newtheorem{remark}{Remark}
\newtheorem{assumption}{Assumption}

\newtheorem{problem}{Problem}
\newtheorem{corollary}{Corollary}

\DeclareMathOperator{\diag}{diag}
\DeclareMathOperator{\SOL}{SOL}
\DeclareMathOperator{\rank}{rank}

\DeclareMathOperator{\supp}{supp}
\DeclareMathOperator{\BR}{BR}
\usepackage{xcolor}

\title{\LARGE \bf
Bayesian hypergame approach to equilibrium stability and robustness
in moving target defense
}

\author{Hanzheng Zhang,  Zhaoyang Cheng, Guanpu Chen,  Karl Henrik Johansson
\thanks{*This work was supported by the Natural Science Foundation of China under Grant 72171171, and also supported by the Swedish Research Council Distinguished Professor Grant 2017-01078, by the Knut and Alice Wallenberg Foundation Wallenberg Scholar Grant, and by the Swedish Strategic Research Foundation SUCCESS Grant FUS21-0026. 
\textit{(Corresponding author:  Guanpu Chen)}
}
\thanks{H. Zhang is with Shanghai Research Institute for Autonomous Intelligent Systems, Tongji University, Shanghai, China.
		{\tt\small zhanghanzheng@tongji.edu.cn}}%
\thanks{Z. Cheng is with Academy of Mathematics and System Science, Chinese Academy of Science, and University of Chinese Academy of Science, Beijing, China.
{\tt\small chengzhaoyang@amss.ac.cn}}%
\thanks{G. Chen  and K. H. Johansson are with the Division of Decision and Control Systems, School of Electrical Engineering and Computer Science, KTH Royal Institute of Technology, and the Digital Futures, Stockholm, Sweden.
        {\tt\small guanpu@kth.se, kallej@kth.se}}
}

\begin{document}

\maketitle
\thispagestyle{empty}
\pagestyle{empty}

\begin{abstract}

We investigate the equilibrium stability and robustness in a class of moving target defense problems, in which players have both incomplete information
and asymmetric cognition. We first establish a Bayesian Stackelberg game model for incomplete information and then employ a hypergame reformulation to address asymmetric cognition. With the core concept of the hyper Bayesian Nash equilibrium (HBNE), a condition for achieving both the strategic and cognitive stability in equilibria can be realized by
solving linear equations. 
Moreover, to deal with players' underlying perturbed knowledge, we study the equilibrium robustness by presenting a condition of robust HBNE under the given configuration. Experiments evaluate our theoretical results.


\end{abstract}

\section{INTRODUCTION}


In recent years, moving target defense (MTD) has become increasingly important due to its broad applications in cyber-physical systems (CPS) and infrastructure protection \cite{umsonst2020nash,niu2018,umsonst2023,wahab2019}. To investigate the deployment of the defense, Stackelberg-game-based methods are extensively employed in leader-follower MTD problems, where the defender moves first and the attacker determines its strategy after observing the defender's decision \cite{mohammad2019,umsonst2018,jia2016,korzhyk2011}.


Game theory is always a powerful tool in cybersecurity \cite{korzhyk2011,amin2019preface,chen2024approach}. Considering widespread incomplete information,
Bayesian Stackelberg games (BSG) offer an effective framework for modeling uncertainties with extensive applications in CPS, cybersecurity, and network communications \cite{umsonst2023,wahab2019}. In BSG, uncertainties are typically characterized by players' types, while the distribution of types is public knowledge \cite{harsanyi1967,andrey2016,chen2020learning,zhang2023}. Then, players' decision-making is based on expected utility outcomes. 
Nonetheless,  most Bayesian models assume that the player's type corresponds to the actual configuration, regardless of whether the players accurately perceive it. The settings fail to capture the situation when
players have asymmetric cognition of the configuration, which often arises in practice due to external disturbances, bounded rationality, or inadvertent errors \cite{jia2016,cheng2022,Gharesifard2010}.
Due to the underlying different perceptions about the game configuration, players may have asymmetric cognition in practice \cite{cheng2022,bennett1977toward}.
The hypergame approach is an effective model to break down this complex scenario by decomposing it into multiple subjective sub-games \cite{bennett1977toward}.
Through subjective sub-games, players can exploit asymmetric information to achieve better outcomes. Hence, hypergame approaches are extensively employed
in economics and CPS \cite{gharesifard2011evolution,Gharesifard2014,Abhishek2019,Bakker2020}.
Due to asymmetric cognition, players are likely to be suspicious about their acquired knowledge if the rivals' strategies are not as expected, which may lead to the collapse of strategic balance, or even to the breakdown of the system. Hence, the concept of hyper Nash equilibrium (HNE) is important in hypergames \cite{cheng2022,bennett1977toward,sasaki2008}. 
When an HNE is achieved, the players not only avoid changing their strategies unilaterally, but also maintain confidence in their cognition.

However, in practice, complex scenarios inevitably 
involve both incomplete information and asymmetric cognition.
Although the aforementioned works in Bayesian games and hypergames succeed in modeling incomplete information and asymmetric cognition, respectively, these works predominantly focused on only one of these two aspects.

Consider a scenario in MTD problems where an attacker receives vague configuration details, representing a form of incomplete information, whereas a defender knows the exact configuration and is aware of the attacker’s incomplete information.
Such incomplete information and asymmetric cognition prompt us to investigate both the strategic and cognitive stability of an obtained equilibrium. 
Otherwise, the players may alter their strategies since the rival's strategies do not coincide with their expectations. 
Moreover, players' knowledge may also be perturbed by fickle environmental conditions or malicious attacks \cite{umsonst2023,jia2016,berady2021ttp,nour2023survey}, and the robustness of an obtained equilibrium is possibly broken. It is thus necessary to ensure the equilibrium robustness against perturbation. The above two problems related to stability and robustness, arising from the complex interaction of incomplete information and asymmetric cognition, necessitate a unified analytical approach.


In this paper, we investigate the equilibrium stability and robustness in a class of leader-follower MTD problems, where the complex circumstances incorporate both incomplete information and asymmetric cognition. The attacker fails to perceive the exact configuration of the game, but characterizes the incomplete information by a probability distribution. Meanwhile, the defender, aware of the asymmetric cognition, knows the exact configuration and the distribution of the attacker's perceptions. 

In this view, the main contributions are as follows.



\begin{itemize}
\item 
We use a hyper Bayesian Stackelberg game approach to model a class of MTD problems with both incomplete information and asymmetric cognition. Through the concept of the hyper Bayesian Nash equilibrium (HBNE), we provide the analysis of both the strategic and cognitive stability, as well as the robustness.


\item 
We propose a stability condition by solving linear equations to reveal when an equilibrium can become an HBNE (Theorem \ref{thm_stable}). We also provide corollaries based on several typical probability distributions, such as the uniform, Bernoulli, and one-point distributions.

\item
To further handle the underlying perturbation in the leader's observation, we study the equilibrium robustness by the concept of robust HBNE strategies. If the defender's perturbed distribution is close enough to the accurate one, an HBNE indicates a robust HBNE strategy under the given perturbation (Theorem \ref{thm_sensi}).

\end{itemize}


\textbf{Notations:} Let $\mathbb{R}$ be the real Euclidean space, and $\mathbb{R}_+=\{x\in\mathbb{R}:\,x\geq0\}$. For $x=(x_1,\dots,x_n)\in\mathbb{R}^n$, $\supp(x)=\{i\in\{1,\dots,n\}:\,x_i>0\}$ is the support set of $x$, $\diag(x)$ is a $n\times n$ matrix with the diagonal elements equal $x$, and $\chi(x):\mathbb{R}^n\to\mathbb{R}^n$ is the indicative function that the $k$-th entry of $\chi(x)$ equals 1 if $x_k\ne 0$ and equals 0 otherwise. 
$1_m$ and $0_m$ represent $m$-dim vectors with all entries equal to 1 and 0, respectively.

\section{PROBLEM FORMULATION}\label{sec_form}
In this section, we first formulate the leader-follower MTD problem with Bayesian games due to incomplete information.
We then reformulate the problem into a hypergame to study asymmetric cognition. 
We show the problems to solve for the equilibrium stability and robustness under both incomplete information and asymmetric cognition.

\subsection{Bayesian model for incomplete information}

Consider an MTD problem with a defender, a follower, and $K$ targets. The attacker aims to attack the vulnerable targets, while the defender tries to protect them. 
In the MTD setting, the defender and the attacker have $R_d$ and $R_a$ resources to protect and attack the targets, respectively. The defender chooses a strategy to assign the resource $x^k$ on each target $k$.
Thus, the defender's strategy set is
$\Omega_d=\{x\in\mathbb{R}_+^{K}:\,\sum_{k=1}^{K}x^k=R_d\}.$
Usually, when $R_d=1$, the defender deploys its defense with the probability $x^k$ to defend target $k$. Similarly, the attacker assigns resource $y^k$ to attack target $k$, and its strategy set is $\Omega_a=\{y\in\mathbb{R}_+^{K}:\,\sum_{k=1}^Ky^k=R_a\}.$ The defender and the attacker maximize their utilities:
\begin{equation}\label{eq_actual_utility_d}
U_d(x, y)=\sum_{k=1}^K y^k(x^k U_d^c(t_k)+(R_d-x^k)U_d^u(t_k)),\end{equation}
\begin{equation}\label{eq_actual_utility_a}
U_a(x,y)=\sum_{k=1}^{K} y^k(x^k U_a^c(t_k)+(R_d-x^k)U_a^u(t_k)).
\end{equation}
Specifically, $U_d^c(t_k)$ and $U_a^c(t_k)$ are the defender's and attacker's profits when the defender allocates per unit of resource to target $t_k$ against per unit attack from the attacker, respectively. The term $R_d-x^k$ refers to the action of not being protected. It is necessary to account for the corresponding profits of not being protected since different profits may be achieved when the attacker attacks different unprotected targets. Similarly, $U_d^u(t_k)$ and $U_a^u(t_k)$ are the defender's and attacker's profit when the defender fails to defend target $t_k$ against per unit attacking resource, respectively.

Due to widespread uncertainties, the attacker may fail to perceive the exact configuration of the MTD problem and thus have incomplete information \cite{umsonst2023,jia2016,chen2021distributed}. The MTD problem can correspondingly be formulated by a two-player BSG $G(\theta,P)=\{\Omega_d\times\Omega_a,\theta,P,\{U_a\}\cup \{U_d\}\}$. Specifically, the defender, with the strategy set $\Omega_d$, is a leader acting first in the Stackelberg game setting, while the attacker, with the strategy set $\Omega_a$, is a follower determining the strategy after observing the leader's strategy. The type $\theta$, on behalf of the attacker's incomplete information of the configuration, is a discrete random variable over the type set $\Theta=\{\theta_1,\dots,\theta_n\}$ from a public distribution $P$. The attacker obtains one type each time, while the defender does not know the attacker's exact type $\theta$, but can get the public distribution $P$ \cite{harsanyi1967}.

When obtaining a type $\theta\in\Theta$, the attacker chooses a strategy $ y(\theta)$ from $\Omega_a$, and assigns the resources $ y^k(\theta)$ to target $k$. In the rest of this paper, we use $y$ to denote a map from $\Theta$ to $(\Omega_a)^n$ and $y(\theta)$ to represent an action in $\Omega_a$ under type $\theta$. In this view, the utility of the defender is still the same as \eqref{eq_actual_utility_d}, while the attacker's utility should be considered under type $\theta$, that is,
$$U_a(x,y,\theta)=\sum_{k=1}^K y^k(\theta)(x^k U_a^c(t_k,\theta)+(R_d-x^k)U_a^u(t_k,\theta)).$$




With a given type $\theta\in\Theta$, the attacker's best response to the defender's strategy $x$ should be
$$\BR(x,\theta)=\arg\max_{y\in\Omega_a}U_a(x,y,\theta).$$

Due to the attacker's incomplete information, the defender makes decisions under this Bayesian frame to maximize the following expectation
$$\begin{aligned}
EU_d(x,\! y)\!=\!\!\sum_{k=1}^K \!\sum_{\theta\in\Theta}\!\! P(\theta) y^k(\theta) (x^k U_d^c(t_k)\!+\!(R_d\!-\!x^k)U_d^u(t_k)\!).
\end{aligned}$$
Hence, we give the following definitions characterizing an equilibrium concept.

\begin{definition}
A strategy pair $(x^\ast, y^\ast)$ is a Bayesian strong Stackelberg equilibrium (BSSE) if 
$$(x^\ast, y^\ast)\in\arg\max_{x\in\Omega_d,\atop  y(\theta)\in \BR(x,\theta)} EU_d(x, y).$$
\end{definition}

The concept of BSSE characterizes strategic stability in $G(\theta,P)$. The defender acts first by making decisions based on the anticipation of the attacker's best response to maximize the expectation of the utility, while the attacker has to employ the best response strategy for its maximal utility. 

In addition to BSSE, the Bayesian Nash equilibrium (BNE) can be reached when both players make decisions as the best response to the opponent's strategy. When reaching a BNE, no player can unilaterally change their strategies for a better payoff. Technically, we give the definition as follows.

\begin{definition}
A strategy pair $(x^\ast,y^\ast)$ is a Bayesian Nash equilibrium (BNE) if 
$$
\begin{aligned}
x^\ast&\in\arg\max_{x\in\Omega_d}\sum_{\theta\in\Theta}P(\theta) U_d(x, y^\ast(\theta)),\\
 y^\ast(\theta)&\in\arg\max_{y(\theta)\in\Omega_a}U_a(x^\ast,y,\theta),\ \ \ \ \forall\, \theta\in\Theta.\end{aligned}$$

\end{definition}


\subsection{Hypergame reformulation for asymmetric cognition}

Consider that, besides the attacker's incomplete information in $G(\theta,P)$, the defender can also realize the exact configuration $\theta_0$ of the game.
Let us denote the MTD problem without incomplete information by $G(\theta_0)$, that is, $G(\theta_0)=G(\theta,P)$ when $\theta\equiv \theta_0$ and $P(\theta_0)\equiv 1$. In this formulation, the attacker is involved in the Bayesian game $G(\theta,P)$, while the defender is involved in $G(\theta_0)$
and also realizes $G(\theta,P)$ of the attacker. 
Thus, the defender should sufficiently utilize this asymmetric cognition
to achieve an improved utility.

We reformulate this setting by a hypergame \cite{bennett1977toward},
which decomposes complex situations with asymmetric cognition into multiple subjective sub-games. 
Take a Bayesian hypergame model $H(P)=\{H_d(P),H_a(P)\}$, where $H_d(P)=\{G(\theta_0),G(\theta,P)\}$ and $H_a(P)=\{G(\theta,P),G(\theta,P)\}$ are the two subjective sub-games of the defender and the attacker, respectively. From the attacker's perspective $H_a(P)$, the attacker believes that both two players are involved in the BSG $G(\theta,P)$ and can obtain the type $\theta$ in each iteration. From the defender's perspective $H_d(P)$, the defender knows not only $G(\theta_0)$ with the exact configuration $\theta_0$ but the attacker's public distribution $P$ in the subjective sub-game $G(\theta,P)$ as well. 
Actually, $H(P)$ is a second-level hypergame\footnote{
The first-level hypergame describes the situation when players are playing different sub-games but no one realizes this. The second-level hypergame occurs when at least one player is aware of the difference between their sub-games. More details can be found in \cite{bennett1977toward}.} and can be summarized in Table \ref{table1}.




\begin{table}[!th]
\renewcommand\arraystretch{1.5}
\caption{Bayesian hypergame $H(P)$}
\vspace{-10 pt}
\label{table1}
\begin{center}
\begin{tabular}{|cc|cc|}
\hline
\multicolumn{4}{|c|}{Bayesian hypergame $H(P)$}\\
\hline
\multicolumn{2}{|c|}{defender $H_d(P)$} & \multicolumn{2}{c|}{attacker $H_{a}(P)$}\\
\hline
defender & attacker & defender & attacker\\
$G(\theta_0)$ & $G(\theta,P)$ & $G(\theta,P)$ & $G(\theta,P)$\\
\hline
\end{tabular}
\end{center}
\vspace{-15 pt}
\end{table}

Due to asymmetric cognition, players may realize that the rivals are involved in a different game when the strategies of the rivals are not consistent with the expected ones according to their subjective sub-games. Then, players may change their cognition and choose other strategies accordingly, leading to the ruin of balance. To this end, we introduce the following important concept in this hypergame reformulation. 

\begin{definition}
A strategy pair $(x^\ast,  y^\ast)$ is a hyper Bayesian Nash equilibrium (HBNE) of $H(P)$ if 
$$
\begin{aligned}
x^\ast&\in\arg\max_{x\in\Omega_d}\sum_{\theta\in\Theta}P(\theta) U_d(x, y^\ast(\theta),\theta_0),\\
 y^\ast(\theta)&\in\arg\max_{y(\theta)\in\Omega_a}U_a(x^\ast,y,\theta),\ \ \ \ \forall\, \theta\in\Theta.\end{aligned}$$
\end{definition}

An HBNE can ensure a balance where the strategies of the rivals are consistent with their own expectations. Thus, when an HBNE is reached, both players are unwilling to change their cognition, leading to the cognitive stability of an equilibrium.

\begin{remark}
There are researches for various situations with players' different cognition via Bayesian games, such as asymmetric posterior belief in sequential games \cite{fudenberg1991perfect,basar1998} and the unawareness of parts of configurations in normal-form games \cite{meier2014bayesian,halpern2006extensive}.
However, in their models still with the same prior knowledge of the configuration, players may update their posterior cognition after observing the rivals' strategies due to asymmetric knowledge. Different from these works, our model concerns the prior asymmetric cognition, where the defender can realize both the real configuration and the attacker's incomplete prior knowledge.
\end{remark}

\subsection{Problem statement}



Under the framework of the BSG model and its hypergame reformulation, we examine the stability of equilibrium points reached by the players. By definition, the equilibrium is strategically stable when players achieve a BSSE. This ensures that the players have no incentive to deviate from their strategies, namely, neither the leader nor the follower would unilaterally alter their strategies to gain a better payoff. 

Besides, when an equilibrium meets the criteria of an HBNE, it corresponds to the optimal solution for each player's subjective sub-game. This indicates cognitive stability, ensuring that players trust the game structure and are unaware of differences between their subjective perceptions of the game and those of their opponent. Without cognitive stability, there is a risk that players might recognize a misalignment, which could ultimately disrupt the balance or even ruin the game mechanism.

Therefore, it is crucial to achieve an equilibrium that has both the strategic and cognitive stability. 
We summarize this aim as the following problem.

\begin{problem}\label{pf_1}
When is an equilibrium both strategically and cognitively stable, that is, when a BSSE is also an HBNE?

\end{problem}

On the other hand, the defender's knowledge often relies on external methods and historical information \cite{umsonst2023,jia2016,zhang2023}. This process can be influenced by rivals and the environment, especially malicious attacks \cite{nour2023survey,berady2021ttp}. 
Under this circumstance, the defender may perceive a perturbed distribution $P'$ of the distribution $P$. Although knowing the exact configuration $G(\theta_0)$, the defender believes that the attacker is involved in $G(\theta,P')$, while the attacker truly engages in  $G(\theta,P)$. 
The hypergame $H(P,P')=\{H_d(P'),H_a(P)\}$ with the defender's perturbed knowledge is summarized in Table \ref{table2}.

\begin{table}[h]
\renewcommand\arraystretch{1.5}
\caption{Perturbed Bayesian hypergame $H(P,P')$}
\label{table2}
\vspace{-10 pt}
\begin{center}
\begin{tabular}{|cc|cc|}
\hline
\multicolumn{4}{|c|}{Perturbed Bayesian hypergame $H(P,P')$}\\
\hline
\multicolumn{2}{|c|}{defender $H_d(P')$} & \multicolumn{2}{c|}{attacker $H_{a}(P)$}\\
\hline
defender & attacker & defender & attacker\\
$G(\theta_0)$ & $G(\theta, P')$ & $G(\theta,P)$ & $G(\theta,P)$\\
\hline
\end{tabular}
\end{center}
\vspace{-15 pt}
\end{table}

Thus, the defender's HBNE strategy in $G(\theta,P')$ is possibly not consistent with the expectation of the attacker. Consequently, our goal is to investigate the following problem for equilibrium robustness under a perturbed distribution $P'$.

\begin{problem}\label{pf_2}
When is the defender's strategy within HBNE of $H(P)$ robust to the perturbed distribution $P'$ of $H(P,P')$?
\end{problem}

The following assumptions about the utilities are essential, and are widely employed in Stackelberg games and MTD problems \cite{umsonst2023,korzhyk2011,cheng2022}.
\begin{assumption}\label{ass_1}
For $k=1,\dots,K$ and $\theta\in\Theta$,
\begin{enumerate}\renewcommand\labelenumi{(\roman{enumi})}
\item $\Delta U_d(t_k)=U_d^c(t_k)-U_d^u(t_k)>0$.
\item $\Delta U_{a}(\theta,t_k)=U_{a}^u(\theta,t_k)-U_{a}^c(\theta,t_k)>0$.
\end{enumerate}
\end{assumption}
Assumption \ref{ass_1}(i) ensures that the defender can benefit from defending a target, while Assumption \ref{ass_1}(ii) indicates that the attacker tends to attack these unprotected targets.

\section{STABILITY ANALYSIS}\label{sec_stable}

In this section, we provide the stability condition when a BSSE is an HBNE, that is, to solve Problem \ref{pf_1}.
We first show the existence of BSSE and HBNE.

\begin{lemma}\label{lem_ex}
Under Assumption 1, there exists a BSSE and an HBNE.
\end{lemma}

We give some interpretation of its proof.  
We adopt the Harsanyi transformation \cite{harsanyi1967} to convert the Bayesian game into a complete-information one.
We introduce a virtual player called `nature', who acts first and determines the type $\theta$ according to its probability $P$.
Consider a Stackelberg game $\tilde{G}$, where both players make decisions after observing nature's action. Hence, in $\tilde{G}$,
the leader has the action set $\Omega_d$ while the follower can adopt an $nK$-dimensional strategy $\tilde{y}=(\tilde y_j^k)$ ($j\in\{1,\dots,n\},k\in\{1,\dots,K\}$) with the action set $\tilde \Omega_{a}=\{\tilde y\in(\Omega_{a})^n:\,\sum_{k=1}^K \tilde y_{j}^k=R_a, \tilde y_{j}^k\geq0\}$. 
In this way, the BSG $G(\theta,P)$ is a complete-information game, where the utilities of the defender and attacker are denoted by $\tilde U_d$ and $\tilde{U}_a$, respectively. Specifically,
$\tilde U_d(x,\tilde{y})=\sum_{j=1}^n\sum_{k=1}^K P(\theta_{j}) U_d(x,(\tilde y_{j}^k)(\theta_j),\theta_0),$ while $
\tilde U_a(x,\tilde{y})=\sum_{j=1}^n\sum_{k=1}^K P(\theta_j) U_a(x,\tilde y_{j}^k,\theta_j).
$

By the transformation, the BSG $G(\theta,P)$ is equivalent to the above complete-information game $\tilde{G}$. Thus, with the results regarding the Nash equilibrium and strong Stackelberg equilibrium in finite normal-form games \cite{basar1998}, the existence of the BSSE and the HBNE is correspondingly achieved.

Take $(x_{BSSE}, y_{BSSE})$ as a pair of BSSE. We provide the stability condition when a BSSE becomes an HBNE, namely, to address Problem 1. 
At first, let us define the following solution set to the linear equations of utilities and types:
\begin{equation*}
\begin{aligned}
\SOL( y)&=\{{y}'\in({\Omega}_a)^n,\lambda>0:\, A_1  y'=\lambda B  y, A_2  y'=0\},\\
A_1&=(A(\theta_1),\dots,A(\theta_n)),\\
A(\theta_j)&=P(\theta_j)\diag\left(\frac{\Delta U_a(\theta_j,t_1)}{\Delta U_d(t_1)},\dots,\frac{\Delta U_a(\theta_j,t_K)}{\Delta U_d(t_K)}\right),\\
B&=(P(\theta_1)I_{K},P(\theta_2)I_{K},\dots,P(\theta_n)I_{K}),\\
A_2&=\diag(1_{nK}-\chi( y)).
\end{aligned}
\end{equation*} 

\begin{theorem}\label{thm_stable}
Under Assumption \ref{ass_1}, if $\SOL(y_{BSSE})$ is nonempty, then the BSSE is also an HBNE in $H(P)$.

\end{theorem}

Nonempty $\SOL(y_{BSSE})$ indicates that $y_{BSSE}$ corresponds to a point of the set $(\Omega_a)^n$, which can be seen via a deterministic linear transformation composed from the parameters of players' utilities and types. Once it is satisfied, both the defender and the attacker will trust their cognition and not change their strategies. The attacker is not able to find that the defender notices the exact configuration $\theta_0$, and will not change its strategy. Besides, the defender will trust its strategy to bring not only cognitive stability but also strategic stability.

Based on Theorem \ref{thm_stable}, we give corollaries for three typical cases. 

\noindent\textbf{Case 1.} Consider uniform distribution $P_{U}$ over $n$ types $\theta_1,\dots,\theta_n$. It is a common scenario in practice and can also cover the single-leader-multi-followers Stackelberg models \cite{cheng2022} by considering the followers' type as individual players. Define 
\begin{equation*}
\begin{aligned}
\SOL^\circ(  y)&=\{{y}^\circ\in({\Omega}_a)^n,\lambda>0:\, A_1^\circ y^\circ\!=\!\lambda B^\circ y, A_2^\circ y^\circ\!=0\},\\
A_1^\circ&=(A^\circ(\theta_1),\dots,A^\circ(\theta_n)),\\
A^\circ(\theta_j)&=\diag\left(\frac{\Delta U_a(\theta_j,t_1)}{\Delta U_d(t_1)},\dots,\frac{\Delta U_a(\theta_j,t_K)}{\Delta U_d(t_K)}\right),\\
B^\circ&=(I_{K},\dots,I_{K}),\\
A^\circ_2&=\diag(1_{nK}-\chi( y)).
\end{aligned}
\end{equation*}
It is not hard to find that verifying the solution to $\SOL^\circ(  y)$ is much easier than $\SOL(  y)$, and we have the following result.
\begin{corollary}\label{coro_uniform}
Under Assumption \ref{ass_1} with uniform distribution $P_U$, a BSSE is an HBNE if $\SOL^\circ( y_{BSSE})$ is nonempty.
\end{corollary}

\noindent\textbf{Case 2}. Consider Bernoulli distribution $P_B$ regarding the two types $\theta_1,\theta_2$. This setting is often employed to characterize players with two behavioral patterns \cite{umsonst2023,andrey2016} or to describe the reliability of perceptions about adversarial utilities \cite{jia2016}. Suppose $R_d=R_a=1$ and $P_B(\theta_1)\ne0,1$. Otherwise, it degenerates to a one-point distribution.
We provide the following result.

\begin{corollary}\label{coro:single:bern}
Under Assumption \ref{ass_1} and given Bernoulli distribution $P_B$, a BSSE becomes an HBNE if 
$\rank  A< 2K-1$, where
$ A=((A_1')^T,(A_2')^T)^T$ and
\begin{align*}
   A_1'=&(P(\theta_1)A_1'(\theta_1),P(\theta_2)A_1'(\theta_2)) ,\\
   A_1'(\theta_i)=&P(\theta_i)\diag\Big(\chi( y^1(\theta_i))\frac{\Delta U_a(\theta_i,t_1)}{\Delta U_d(t_1)},\\
   &\quad\quad\quad\quad\quad\quad \cdots, \chi( y^K(\theta_i))\frac{\Delta U_a(\theta_i,t_K)}{\Delta U_d(t_K)}\Big),\\
   A_2'=&\diag(1_{2K}-\chi(y)).
\end{align*}
\end{corollary}


\noindent\textbf{Case 3.} Consider one-point distribution $P_O$. In this scenario, the BSG poses a single type $\theta\equiv\theta_0$ and degenerates to a complete-information game. In such a classic leader-follower scheme, the relationship between the Stackelberg and Nash equilibrium has been widely investigated in cybersecurity problems \cite{korzhyk2011,xu2024consistency}, which can also be corroborated by our result. We give the result as follows.
\begin{corollary}
Under Assumption 1 and given one-point distribution $P_O$, a BSSE is an HBNE.


\end{corollary} 

\section{ROBUSTNESS ANALYSIS}\label{sec_sensi}



In this section, we investigate the robustness of HBNE, \textit{i.e.}, to solve Problem \ref{pf_2}.

%


When both players engage in $H(P,P')$, they make decisions according to their subjective sub-games, which are different actually. If the defender's decision-making is not as the attacker expects due to the perturbation $P'\neq P$, the attacker may realize that their cognition of the game configuration is not consistent. This indicates the lack of robustness to $P'$ regarding the defender's strategy and may lead to the ruin of the system. To this end, we introduce the concept of \textit{robust} HBNE strategy in $H(P,P')$ with the defender's perturbed distribution $P'$ \cite{sasaki2008}.
\begin{definition}
Consider HBNE $(x^\ast,y^\ast_P)$ of $H(P)$ and a perturbed distribution $P'\neq P$. Then,
$x^\ast$ is said to be the defender's  \textit{robust} HBNE strategy of $H(P,P')$ if there exists $y^\ast_{P'}(\theta)\in \BR(x^\ast,\theta)$ for $\theta\in\Theta$ such that 
\begin{equation}\label{eq_def_robust_hbne}
x^\ast\in\arg\max_{x\in\Omega_d} \sum_{\theta\in\Theta} P'(\theta) U_d(x,y^\ast_{P'}(\theta),\theta_0).
\end{equation}

\end{definition}

Like the  pair of HBNE $(x^\ast,y^\ast_P)$ in $H(P)$ without perturbation, the pair of $(x^\ast,y^\ast_{P'})$ can ensure both the strategic and cognitive stability in $H(P,P')$. 
Note that $y^\ast_{P'}$ is a virtual strategy since the attacker still acts $y^\ast_P$ in its subjective sub-games of $H(P,P')$.
When the defender acts robust HBNE strategy $x^*$, the attacker cannot realize that the defender has perturbed knowledge. 
This implies the robustness of the defender's strategy $x^*$ from $P$ to $P'$.
This motivates us to reveal the following result.

\begin{theorem}\label{thm_sensi}
Suppose a pair of HBNE $(x^\ast, y^\ast_P)$  in $H(P)$. Under Assumption \ref{ass_1}, for a perturbed distribution $P'$, there exists a positive constant $\epsilon$ determined by players' utilities such that, if $\sum_{\theta\in\Theta} |P'(\theta)-P(\theta)|< \epsilon$, then
$x^\ast$ is a robust HBNE strategy of $H(P,P')$.
\end{theorem}

Theorem \ref{thm_sensi} reveals that, when $P'$ is close to $P$, the attacker cannot realize that the defender's observation has perturbation, and thus will trust its own cognition of the game configuration.
On the other hand, we could say that the defender's HBNE strategy $x^*$ of $H(P)$ is robust to the perturbed distribution in $H(P,P')$.
When the defender observes the distribution of the attacker's type, a small bias is usually allowed according to Theorem \ref{thm_sensi}, which does not influence the defender's decision-making.

\section{NUMERICAL SIMULATIONS}\label{sec_simu}

In this section, we provide numerical simulations to verify our theoretical results.

\subsection{Stable condition for BSSE being HBNE}

\begin{figure}[t]
\centering
\subfloat[$n=5$]{\includegraphics[width=1.78in]{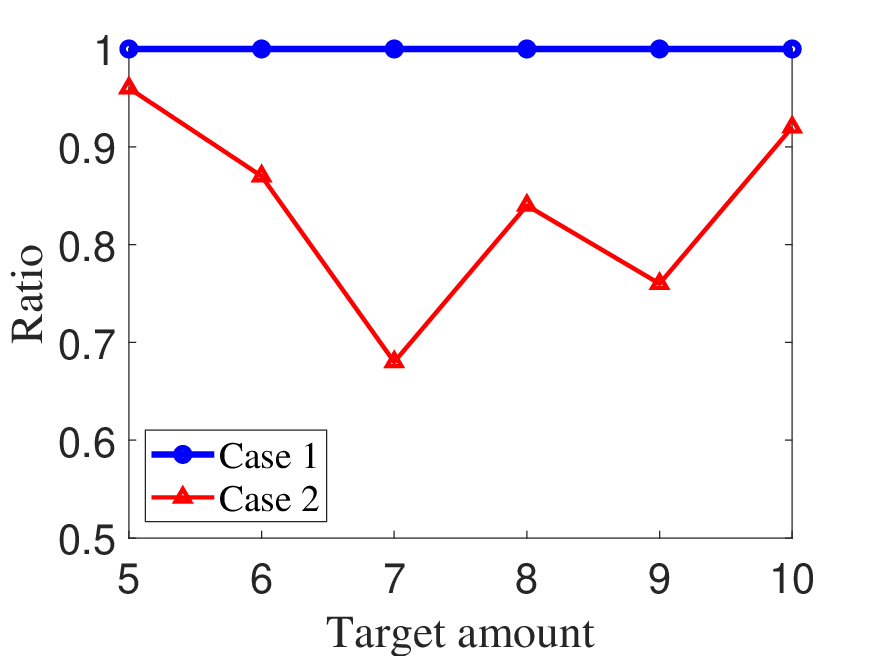}}
\subfloat[$K=5$]{\includegraphics[width=1.78in]{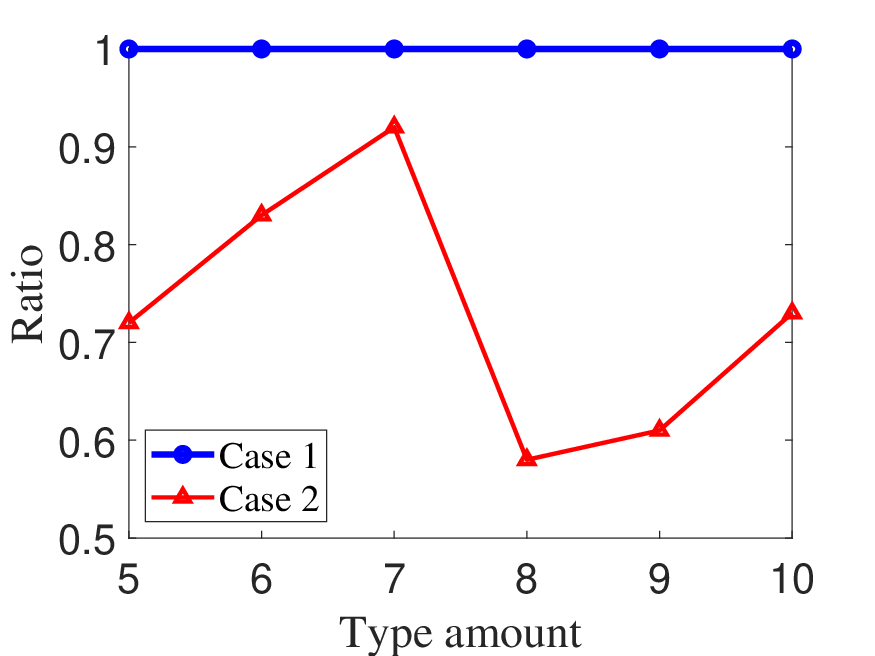}}
\caption{The equivalence of BSSE and HBNE.
The x-axis is for the target amount in (a), and for the type amount in (b). The y-axis is for ratios where Case 1 and Case 2 happen in 100 instances, depicted in blue and red, respectively.}\label{fig_BSSE}
\end{figure}

To evaluate Theorem \ref{thm_stable}, consider a randomly generated defense model, which has $n$ types and $K$ targets with $n=5,\dots,10$ and $K=5,\dots,10$. Generate 100 instances for each model, where $U_d^c(t_k)$ and $U_a^u(\theta,t_k)$ are uniformly generated over $[5,10]$, $U_d^u(t_k)$ and $U_a^c(\theta,t_k)$ are uniformly generated over $[0,5]$, and $R_d$ and $R_a$ are uniformly generated over $[1,5]$. Then, Let $(x_{BSSE},y_{BSSE})$ be the BSSE of each instance, which is computed by the mixed-integer linear programming in \cite{korzhyk2011}, while the HBNE is computed by sequential
quadratic programming in \cite{chatterjee2009optimization}. We investigate the following two cases. Case 1: a BSSE is an HBNE when $\SOL(y_{BSSE})$ is nonempty; Case 2: $\SOL(y_{BSSE})$ is nonempty when a BSSE is an HBNE. Fig. \ref{fig_BSSE} gives the ratios where the two cases happen in all the instances, and we can see that the ratio of Case 1 is always $100\%$, which is consistent with Theorem \ref{thm_stable}. The ratio of Case 2 is no less than $75\%$, which means that the proposed stability condition in Theorem \ref{thm_stable} can cover many instances.

\subsection{Robustness of HBNE}

\begin{figure}[t]
\vspace{-15pt}
\centering
\subfloat[Setting 1]{\includegraphics[width=1.78in]{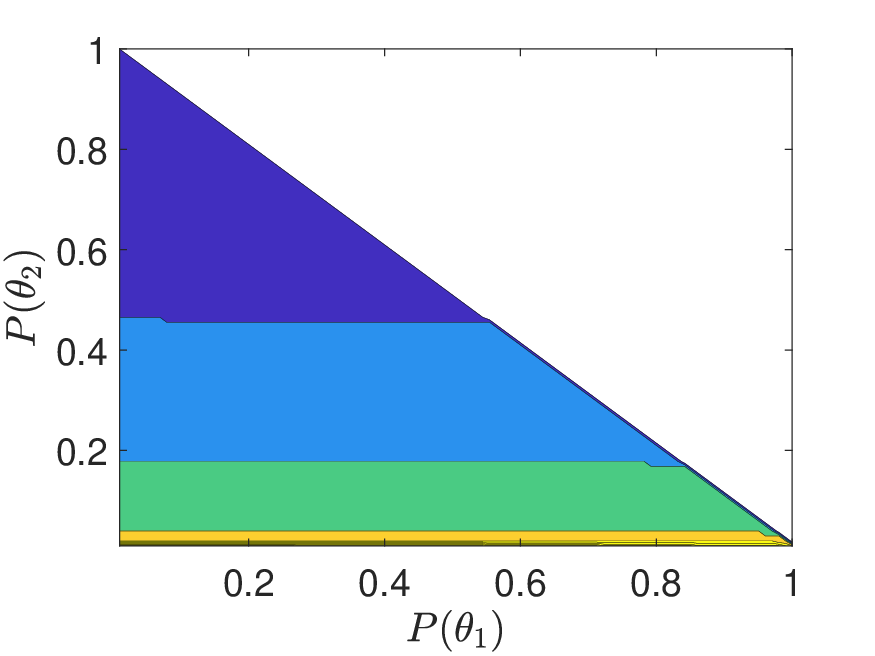}}
\subfloat[Setting 2]{\includegraphics[width=1.78in]{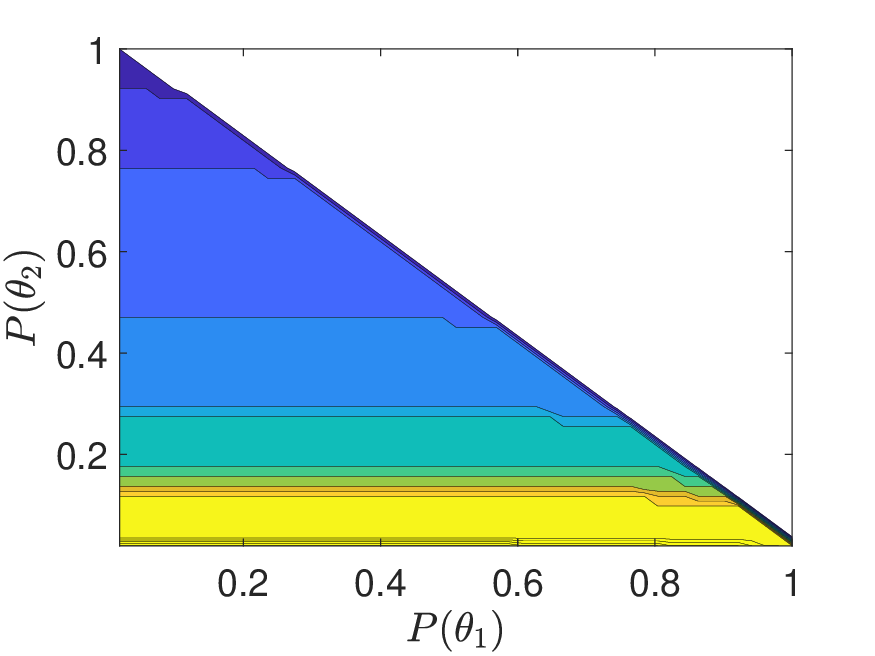}}
\caption{Robust HBNE in two settings. Any point $(P(\theta_1),P(\theta_2))$ in this figure refers to a distribution $P$ over $\theta_1$, $\theta_2$, and $\theta_3$ with $P(\theta_3)=1-P(\theta_1)-P(\theta_2)$. 
If a point has a same-color neighborhood, then the corresponding HBNE is robust. Figures (a) and (b) are two different settings with randomly generated utilities.
}\label{fig_hbne}
\vspace{-10pt}
\end{figure}

Consider an MTD problem against malicious attacks in cloud systems, where the hacker has several attack methods such as ``denial-of-service'', ``malware injection'', and ``side-channel attacks'', while the administrator needs to resist these threats\cite{masdari2016survey,schlenker2018deceiving}. Both sides have limited resources (e.g. budgets, hardware). The hacker cannot obtain complete information on the configuration of the system possibly due to observation error and defensive deception. 
Moreover, the administrator also does not know the hacker's perception (type), and gets a perturbed distribution of the type due to malicious jamming. Consider the hacker's distribution $P$ over 3 types $\theta_1,\theta_2,\theta_3$, and the parameters $U_d^c(t_k)$,$U_a^u(\theta,t_k)$,$U_d^u(t_k)$,$U_a^c(\theta,t_k)\in[0,5]$ as the value for different targets and set $R_d=R_a=1$. {Fig. \ref{fig_hbne} illustrates the defender's HBNE strategies in two given settings, where the parameters are randomly generated.} Each point $(P(\theta_1),P(\theta_2))$ in Fig. \ref{fig_hbne} refers to a distribution with $P(\theta_3)=1-P(\theta_1)-P(\theta_2)$. 
If a point has a same-color neighborhood, then the
corresponding HBNE is robust.
We can see that most points of its distribution are in the same color as its neighborhood. 
Thus, most HBNE strategies are robust to malicious jamming, which is consistent with Theorem \ref{thm_sensi}.

\section{CONCLUSIONS}\label{sec_con}


We studied MTD problems incorporating both incomplete information and asymmetric cognition. We modeled the complex circumstance as a hyper Bayesian game, where the attacker fails to perceive the exact configuration, and the defender realizes both the exact configuration and the distribution of the attacker's perception. By solving linear equations, we provided a stability condition to reveal both the strategic stability and cognitive stability of obtained equilibria. Further, to handle players' underlying perturbed knowledge, we showed the equilibrium robustness through the concept of the robust HBNE.

\bibliographystyle{ieeetr}
\bibliography{root.bib}

\end{document}